# Domain dynamics in nonequilibrium random-field Ising models

S. HRIVŇAK, M. ŽUKOVIČ *

Institute of Physics, Faculty of Sciences, P. J. Šafárik University, Park Angelinum 9, 041 54 Košice, Slovakia

We employ Monte Carlo simulations in order to study dynamics of the magnetization and domain growth processes in the random-field Ising models with uniform and Gaussian random field distributions of varying strengths. Domain sizes are determined directly using the Hoshen-Kopelman algorithm. For either case, both the magnetization and the largest domain growth dynamics are found to follow the power law with generally different exponents, which exponentially decay with the random field strength. Moreover, for relatively small random fields the relaxation is confirmed to comply with different regimes at early and later times. No significant differences were found between the results for the uniform and Gaussian distributions, in accordance with the universality assumption.

## 1. Introduction

The random-field Ising model (RFIM) has been intensively studied since its introduction by Imry and Ma [1]. It is a prototypical model for magnetic systems with quenched disorder, in which competing mechanisms for order and disorder coexist. While the local spin interactions favor ferromagnetic ordering, the random field variations tend to destroy it. This competition drastically affects thermodynamic properties. As a result, for example, the two-dimensional (2D) RFIM has been shown to display no long-range ordering at any temperature [2]. Thus, in the 2D RFIM statistical mechanics of interfaces or domain walls becomes the key question. Unlike in the zero-field Ising model, in the RFIM it is not always possible to shrink the domain walls to reduce the surface energy and the domain walls are said to be pinned by the local fields. Thus, when the domain walls evolution is finished the system remains in a disordered state, albeit the resulting ferromagnetic domains may be very large. This gives rise to multimodality of the free energy surface and the resulting long relaxation times. Dynamical properties of the 2D RFIM were recently studied for random fields with uniform distribution [3].

The purpose of the present work is to study the behavior of the 2D RFIM in the nonequilibrium region with emphasis on the nature of the magnetization and domain growth processes for uniform and Gaussian random field distributions.

## 2. Model and method

The Hamiltonian of the 2D RFIM can be written in the form

$$H = -J \sum_{\langle i,j \rangle} s_i s_j + \sum_i \eta_i s_i , \qquad (1)$$

where $J$ is the coupling constant, conventionally set to unity, $s_i = \pm 1$ and $\eta_i$ represent respectively the Ising spin and the quenched random field on the $i$*th* site, and $<i,j>$ denotes the summation over nearest neighbors. In the present study, $\eta_i$ is drawn from a zero-mean Gaussian and uniform distributions with varying strengths $\eta_0$. The parameter $\eta_0$ is proportional to the standard deviations of the respective distributions so that also their second central moments match.

We perform Monte Carlo simulations of the RFIM on a square lattice of the size 256 × 256 at a reduced temperature $k_B T/|J| = 0.5$. We employ the Metropolis dynamics and apply periodic boundary conditions to eliminate boundary effects. In most similar studies, domain (cluster of spins in the same state) sizes were estimated indirectly from the fluctuations in magnetization. In the present study, for better precision we directly measure them by Hoshen-Kopelman algorithm [4]. In order to improve the accuracy and the quality of the results we have performed 50 independent simulation runs, and the resulting quantities presented below represent the obtained average values.

## 3. Results and discussion

The time (Monte Carlo sweep) evolution of the magnetization curves for the RFIM with both uniform (URF) and Gaussian (GRF) random fields of different strengths $\eta_0$ show similar behavior and thus in Fig. 1(a) we only present the results for GRF. Apparently, the character of the evolution of the magnetization strongly depends on the strength of the disorder. Namely, before the saturation value is reached the magnetization growth can be characterized by the power law behavior with the $\eta_0$-dependent exponent $\beta(\eta_0)$. The dependence of the power law exponent $\beta(\eta_0)$ on the disorder strength $\eta_0$ is shown in Fig. 2. In both URF and GRF it is found to follow the exponential law with similar decay exponents $c_\beta$, given in Fig. 2. Such a decrease of the exponent $\beta$ with $\eta_0$ could be expected. With the increasing strength of disorder $\eta_0$ spin flips are gradually suppressed owing to the presence of the pinning interaction term (the second term in the Hamiltonian) and consequently the domain walls get pinned, leaving the system in a disordered phase. The fact that the URF and GRF exponents take similar values can be ascribed to the universality phenomenon, which assumes the RFIM properties independent on the choice of the RF distribution.

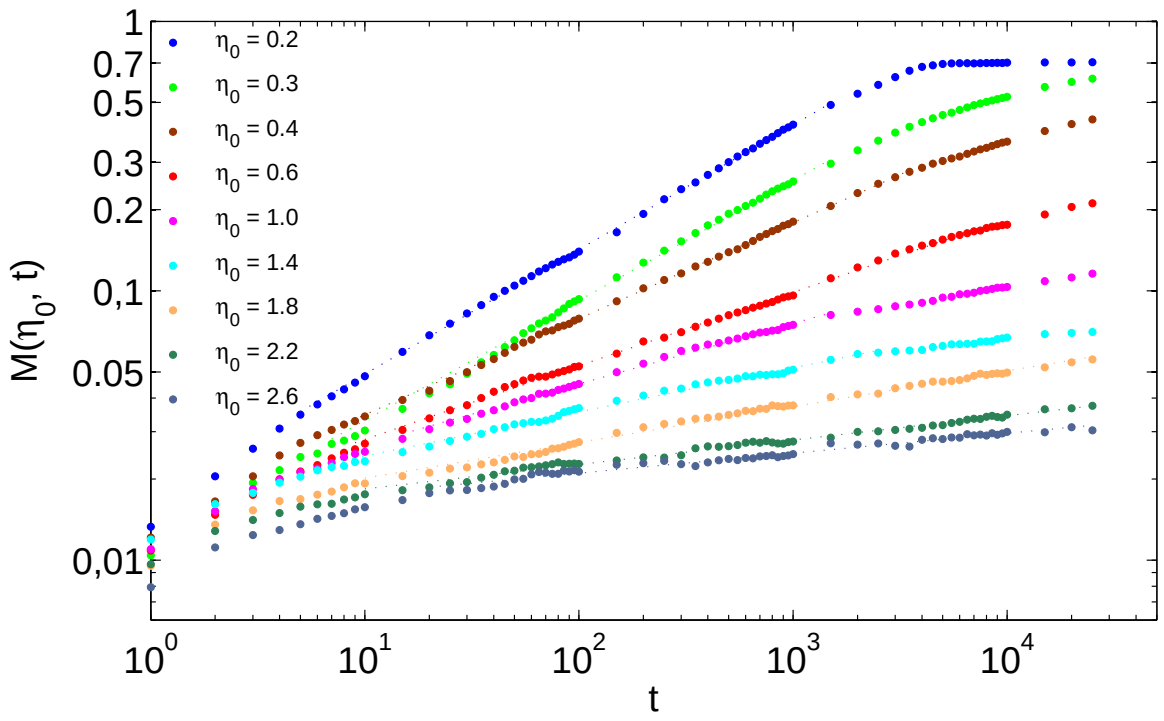

Fig.1. Log-log plots of the magnetization time evolution for GRF with the strengths from $\eta_0$ = 0.2 (top) to $\eta_0$ = 2.6 (bottom). The dashed lines represent the best linear fits.

*corresponding author; e-mail: milan.zukovic@upjs.sk

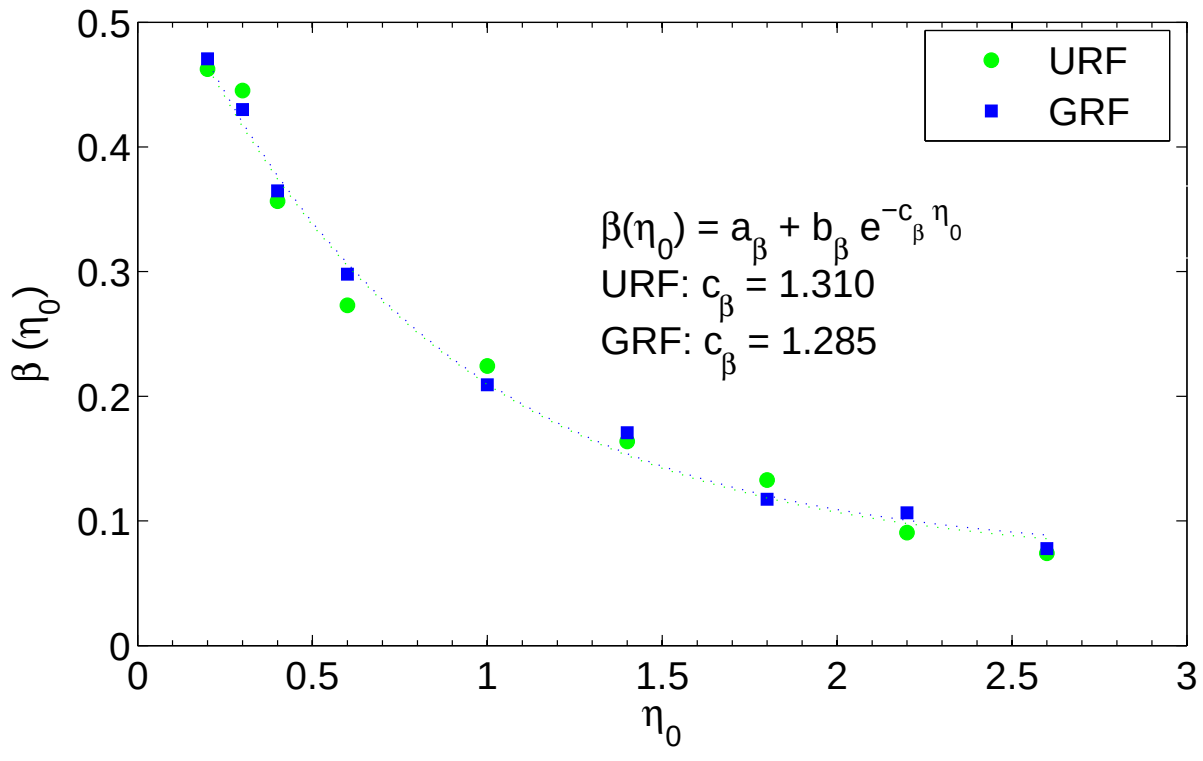

Fig.2. Variations of the magnetization growth exponent $\beta(\eta_0)$ with the RF strength $\eta_0$ for URF and GRF. The dashed lines denote the best exponential fits to the data points.

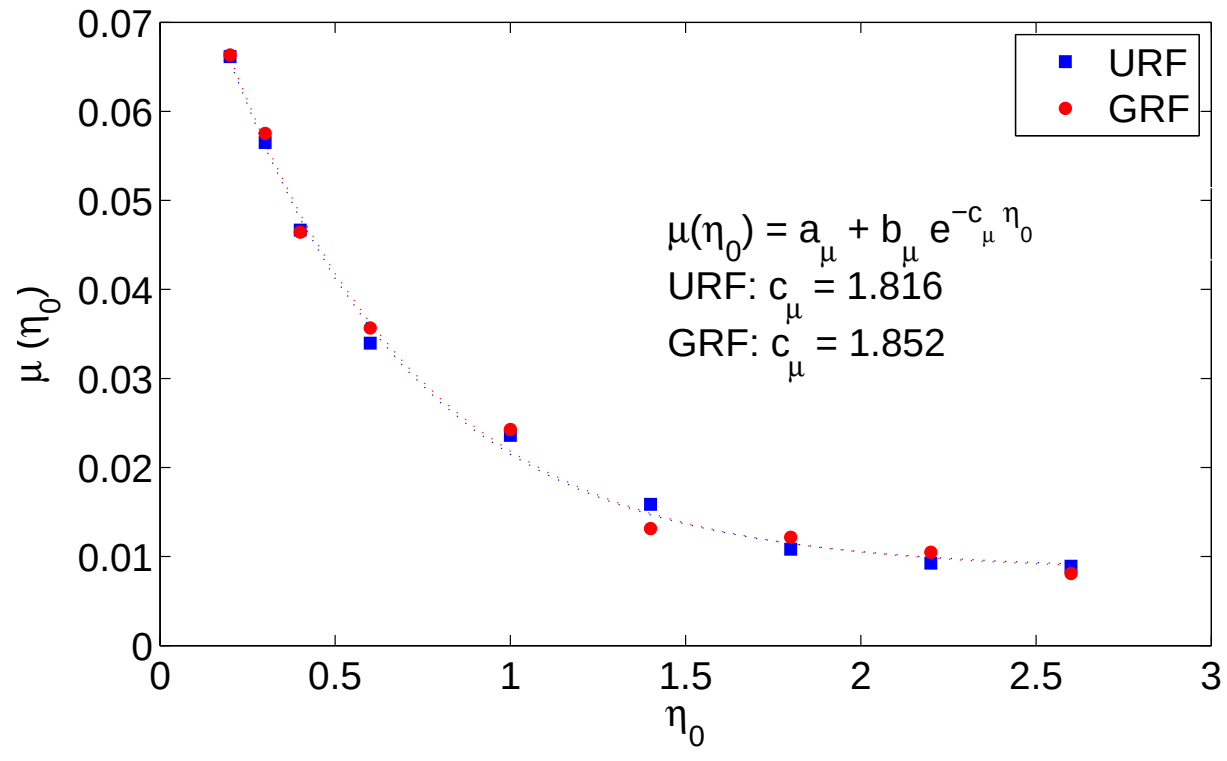

Fig.4. Variations of the largest domain growth exponent $\mu(\eta_0)$ with the RF strength $\eta_0$ for URF and GRF. The dashed lines denote the best exponential fits to the data points.

In Fig.3 we demonstrate the growth dynamics of the largest domain in the GRF case, again for various values of the strength parameter $\eta_0$. For weak disorder ($\eta_0 \leq 0.3$) we observe three distinct time regimes. The early and intermediate time regimes follow the power law behavior with generally different exponents $\mu(\eta_0)$ and $\nu(\eta_0)$, respectively, and the late time regime corresponds to the steady state. For larger disorder strengths ($\eta_0 > 0.3$) there is only one power law regime with the exponent $\mu(\eta_0)$.

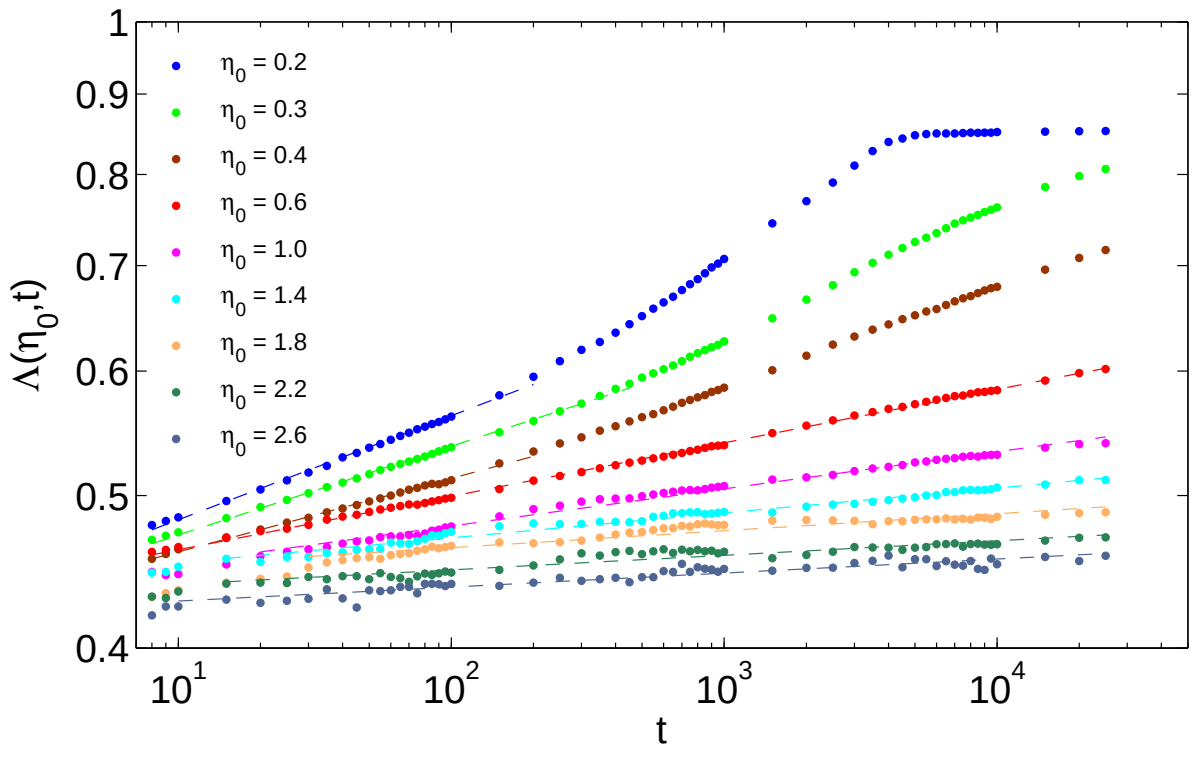

Fig.3. Log-log plots of the largest domain size as a function of time for $\eta_0$ = 0.2 (top) up to $\eta_0$ = 2.6 (bottom). The dashed lines represent the best linear fits to the data points.

The variations of $\mu(\eta_0)$ with $\eta_0$ are plotted in Fig. 4 for both URF and GRF. It is evident that, similar to the magnetization behavior, they fall off exponentially, however, the decay exponents $c_\mu$ are considerably different from the magnetization decay exponents $c_\beta$. Nevertheless, the difference between the URF and GRF values is again very small.

## 4. Conclusions

We studied the nonequilibrium behavior of the 2D RFIM with uniform and Gaussian RF distributions. The dynamic evolution of the magnetization exhibited a power-law growth with the exponent $\beta(\eta_0)$, falling off exponentially with the RF strength $\eta_0$. The growth of the largest domain followed the power law with different exponent $\mu(\eta_0)$, which fell off with $\eta_0$ even faster than $\beta(\eta_0)$. For weak disorder we observed different regimes at early and later times. No significant differences were found between the uniform and Gaussian RF distributions, presumably due to the universality phenomenon.

## Acknowledgement

This work was supported by the Scientific Grant Agency of Ministry of Education of Slovak Republic (Grant No. 1/0234/12). The authors acknowledge the financial support by the ERDF EU (European Union European Regional Development Fund) grant provided under the contract No. ITMS26220120047 (activity 3.2.).

## References

[1] Y. Imry, S.K. Ma, *Phys. Rev. Lett.* **35**, 1399 (1975). DOI: 10.1103/PhysRevLett.35.1399

[2] M. Aizenman, J. Wehr, *Phys. Rev. Lett.* **62**, 2503 (1989). DOI: 10.1103/PhysRevLett.62.2503

[3] S. Sinha, P.K. Mandal, *Phys. Rev. E* **87**, 022121 (2013). DOI: 10.1103/PhysRevE.87.022121

[4] H. Hoshen, R. Kopelman, *Phys. Rev. B* **14**, 3438 (1976). DOI: 10.1103/PhysRevB.14.3438